\documentstyle[psfig,epsf]{article}
\begin{document}
\phantom{anchor}
\vspace*{-1cm}
\begin{flushright} \large
 {\tt University of Bergen, Department of Physics}    \\[2mm]
 {\tt Scientific/Technical Report No.1997-01}    \\[2mm]
 {\tt ISSN 0803-2696} \\[5mm]
 {hep-ph/9701245} \\[5mm]
 {January 1997}           \\
\end{flushright}
\vspace{0.5cm}
\begin{center}
{\Large \bf Probing the $CP$ of the Higgs at an $e^-e^-$ collider}
\footnote{Presented at {\sl QFTHEP '96}, XIth Workshop on
Quantum Field Theory and High Energy Physics, St. Petersburg, Russia, 
September 12-18, 1996, to appear in the Proceedings.}
\\
\vspace{4mm}
Odd Magne Ogreid\\
Department of Physics, University of Bergen, Allegaten 55, N-5007 BERGEN\\
NORWAY\\ 
\end{center}
\begin{abstract}
The sensitivity of various angular and energy distributions to the
$CP$ of a Higgs candidate are considered for the process
$e^-e^-\rightarrow e^-e^-h$, where $h$  is either a scalar or a
pseudoscalar Higgs. This process may differ from the corresponding
production at an $e^+e^-$ collider in the availability of both beams
being polarized. Beam polarization enhances the sensitivity of the
azimuthal distribution.
\end{abstract}
\section{Introduction}
In planning for the Next Linear Collider \cite{Snowmass,ECFA}, it is important
that one in addition to the electron-positron mode also considers the
electron-electron mode \cite{Cuypers}. One reason for this is that one
can produce states not readily accessible in the annihilation channel. Another
reason is that polarization of the electron beams will be readily
available.
The initial state consisting of two electrons is also a
very clean state, there is no background from annihilation processes   

When the Higgs particle is discovered and its mass has been
measured, one will want to explore other properties of this
particle. One of these is whether it is a scalar or a
pseudoscalar. The Higgs predicted by the Standard Model is a scalar
particle. However, certain extensions of the Standard Model allow for
more than one Higgs particle. We can have both scalar and
pseudoscalar Higgs particles in these models. This is the case in the MSSM 
\cite{MSSM}. 
When a possible candidate for the Higgs is
observed, it is therefore of the utmost importance to determine its
properties, and in particular its $CP$.

The possibility of disentangling the
$CP$ of the Higgs particle in the electron-electron mode will here be
considered. This
question has already been considered for the Bjorken
process, where distributions in polar angle provide an easy resolution
\cite{Zerwas}. Also, azimuthal and energy distributions can be used
for an independent determination of the $CP$ \cite{Skjold}.

The tree level diagrams for the process
\begin{eqnarray}
e^-(p_1)+e^-(p_2)\rightarrow e^-(p_1^\prime)+e^-(p_2^\prime)+h(p_h)\nonumber
\end{eqnarray}
are those where the Higgs particle (scalar or
pseudoscalar) is produced via the fusion of two $Z$
particles. There are two diagrams at the tree level because of identical
electrons in the final state. For the $ZZh$ coupling one can take \cite{CKP}
\begin{equation}
i2^{5/4}\sqrt{G_F}\left\{
\begin{array}{ll}
m_Z^2\,g^{\mu\nu} & \mbox{for }h=H \mbox{ ($CP$ even)}, \\
\eta\, \epsilon^{\mu\nu\rho\sigma} k_{1\rho} k_{2\sigma} &
\mbox{for }h=A \mbox{ ($CP$ odd)}.
\end{array}
\right.
\end{equation}
Here, distributions which are obtained when all
outgoing leptons are detected will be considered. The
more favoured production via $WW$ fusion will therefore not be
discussed in this paper. 

One expects the Higgs to be light. Therefore,
in the numerical calculations, a Higgs mass of
135~GeV has been used. This value is consistent with a large number of
models.
\section{Kinematics}
The two electrons in the final
state will be distinguished according to which has the higher energy,
thus $E_1^\prime\geq E_2^\prime$. The angles between the final
state and the initial state electrons are denoted $\theta_1$ and
$\theta_2$, 
\begin{eqnarray}
{\bf p}_1\cdot{\bf p}_1^\prime&=&|{\bf p}_1||{\bf
  p}_1^\prime|\cos\theta_1,
\nonumber\\
{\bf p}_2\cdot{\bf p}_2^\prime&=&|{\bf p}_2||{\bf
  p}_2^\prime|\cos\theta_2.\nonumber 
\end{eqnarray}
Since the process contains three particles in the final state, the
incoming and outgoing particles will not all lie in the same plane.
But the two momenta ${\bf p}_1$ and ${\bf
  p}_1^\prime$ will determine a plane, and similarly for ${\bf p}_2$ and ${\bf
  p}_2^\prime$. The angle between these two planes will be denoted
$\phi$,
\begin{eqnarray}
\cos\phi=\frac{({\bf p}_1\times{\bf p}_1^\prime)
\cdot({\bf p}_2\times{\bf p}_2^\prime)}
{|{\bf p}_1\times{\bf p}_1^\prime|
|{\bf p}_2\times{\bf p}_2^\prime|}.\nonumber
\end{eqnarray}
The degree of longitudinal polarization for each of the initial
electrons will be denoted $P_i$, thus $P_i=+1$ for pure states
of positive helicity and $P_i=-1$ for pure states of
negative helicity.
\begin{figure}[hp]
\centerline{\psfig{figure=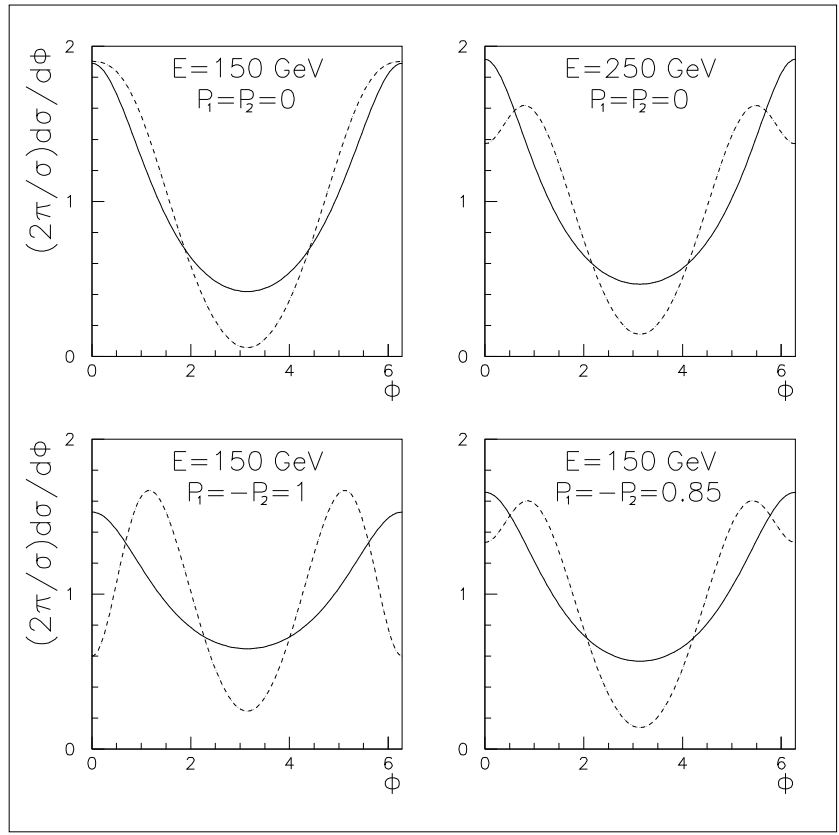,height=11cm,width=11cm}}
\caption{Azimuthal distributions for different values of the beam
  energy and various
  degrees of electron polarizations. The Higgs mass considered is
  135~GeV. The solid lines correspond to a
  scalar Higgs and the dashed lines to a pseudoscalar one.}  
\end{figure}
\section{The cross section}
The four-fold differential cross section
\begin{equation}\label{2dim}
\frac{{\rm d}^4\sigma}{{\rm d}\epsilon\ {\rm d}\phi\ {\rm d}\cos\theta_1\ 
{\rm d}\cos\theta_2}\nonumber
\end{equation}
is calculated in both the scalar and the pseudoscalar case. The
variable $\epsilon$ is the (positive)
energy difference between the two final-state electrons, thus 
$\epsilon=E_1^\prime-E_2^\prime$. This fully differential cross
section is not easily studied experimentally, and therefore only 
cross sections obtained by integrating over two or three
of the variables will be considered. 
In the extreme forward directions the electrons
will be difficult to detect. Therefore, a cut on the polar
angles $\theta_i$ is imposed in such a way that $|\cos\theta_i|\leq0.9$,
whenever an integration over these
variables is performed.
\subsection{Azimuthal distributions}
The azimuthal distribution
\begin{equation}
\frac{2\pi}{\sigma_{\rm cut}}\frac{{\rm d}\sigma_{\rm cut}}{{\rm d}\phi} 
\end{equation}
has been determined by numerical integration, and is shown in
Figure 1 for two different beam energies (150~GeV and 250~GeV) and at various
polarizations for both scalar and pseudoscalar Higgs.  
For the unpolarized case the distributions are rather
similar for a $CP$-odd and a $CP$-even Higgs at a beam energy of
150~GeV. At a higher energy, $E$=250~GeV, one can see a
distinct difference between the two cases. This difference can be
enhanced by turning on electron polarization. In the
case of opposite polarizations (100\%), one sees that the $CP$-even
distribution behaves like $1+\cos\phi$ while the $CP$-odd distribution
behaves more like $1-\cos 2\phi$. The term proportional to
$\cos 2\phi$ has its origin in interference between transverse
$Z$-particles with parallel and orthogonal linear polarizations,
whereas a term proportional to $\cos\phi$ will arise
from interference between scalar and transversely polarized
vector bosons \cite{Ginzburg}.
In the more realistic case of 85\%
electron polarization one can get nearly the same effect as by increasing 
the beam energy to around 250~GeV. 

\subsection{Energy distributions}
From purely kinematical considerations one can show that 
\begin{eqnarray}
0\leq\epsilon\leq\frac{1}{2}E-\frac{1}{8}\frac{m_h^2}{E}
\equiv\epsilon_{\rm max}, \nonumber
\end{eqnarray}
where $m_h$ is the corresponding Higgs mass.
A new variable, the ``scaled energy
difference'', $x=\epsilon/\epsilon_{\rm max}$, $(0\leq x\leq 1)$, is
therefore introduced.
The $x$-distribution
\begin{equation}
\frac{1}{\sigma_{\rm cut}}\frac{{\rm d}\sigma_{\rm cut}}{{\rm d}x}
\end{equation}
is shown in Figure 2. As can be seen from the
plots, the distributions are rather similar in both the $CP$-even and
the $CP$-odd case. A minor
gain can be obtained from polarizing the beams, but probably too small
to be useful in determining the $CP$ of the Higgs.
\vspace{-1cm}
\begin{figure}[h]
\centerline{\psfig{figure=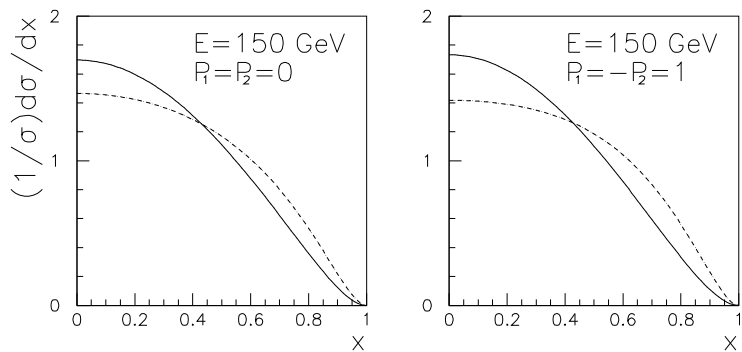,height=5.5cm,width=11cm}}
\caption{The energy difference distribution shown at a 
  beam energy of 150~GeV, and for two different cases of electron polarization.
  For both plots, the Higgs mass considered is 135 GeV. 
  The solid lines correspond to a
  scalar Higgs and the dashed lines to a pseudoscalar one.}  
\end{figure}
\subsection{Polar angle distributions}
Finally, the two-dimensional polar angle distribution
\begin{equation}
\frac{1}{\sigma_{\rm cut}}\frac{{\rm d}^2\sigma}
{{\rm d}\cos\theta_1\ {\rm d}\cos\theta_2},
\end{equation}
is considered.
Representative plots are shown in Figure 3. The plots are made
for the unpolarized case only. Little can be gained by using electron
polarization when considering this distribution.  
The $CP$-odd distribution totally changes nature as one
increases the beam energy from 100~GeV to 200~GeV. The $CP$-odd and
$CP$-even distributions are easier to distinguish from each other at
the lower beam energies. As the beam energy increases, they both peak 
in the forward region where $\cos\theta_1=\cos\theta_2=\pm 1$. But the
production occuring in the forward region for $CP=-1$ is still an
order of magnitude lower than what is the case for $CP=+1$. 
\begin{figure}[ht]
\centerline{\psfig{figure=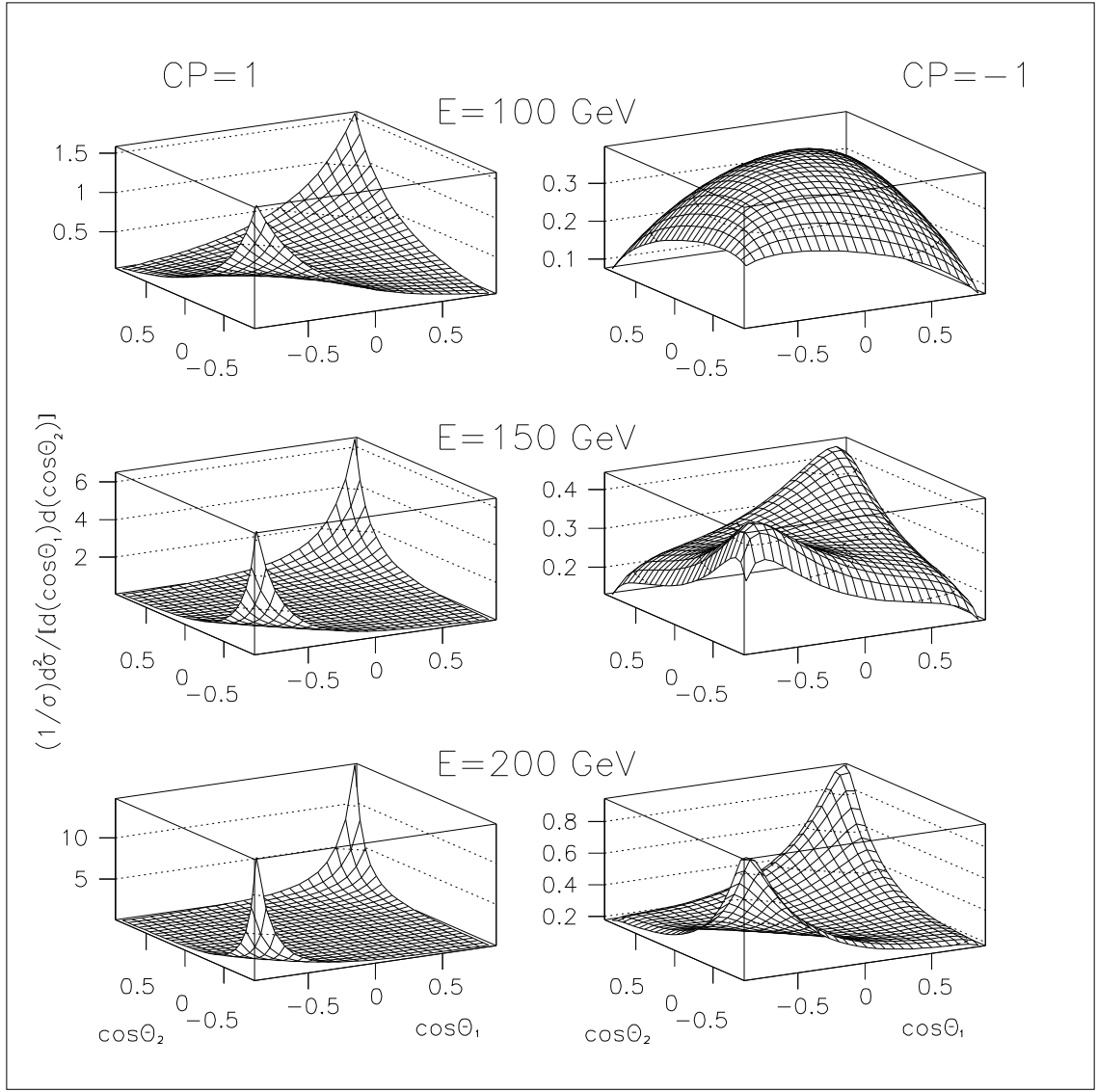,height=15cm,width=15cm}}
\caption{Polar angle distributions for a scalar Higgs
(left figures) and for a pseudoscalar Higgs (right figures) at various
beam energies. For all plots, the Higgs mass considered is 135 GeV.} 
\end{figure}
\section{Summary}
Among the three different distributions presented here,
the azimuthal distribution and the polar angle distribution are those
that contain the clearest signatures of the $CP$ of the
Higgs. The azimuthal distribution can be made quite distinct 
if beam polarization is available, in particular if the two beams are
oppositely polarized. Also, one gets better separation at
higher energies.

For the two-dimensional polar angle distribution, the 
difference between $CP=1$ and $CP=-1$ is most striking at 
lower energies. In the $CP$-even case, the scattered electrons are forward
peaked. As the beam energy increases, the
distributions get more forward peaked. 
In the $CP$-odd case, there is a qualitative change from low to higher
energies. At low energies (or heavy Higgs) the electrons tend to be
scattered out to large angles, $|\cos\theta_i|\simeq0$, whereas at  
higher energies the distributions peak \em near \em the forward direction.
\vspace{0.5cm} 

It is a pleasure to thank my collaborators C.~A.~B\o e and
Per Osland for helpful discussions and suggestions along the
way. I will also thank I.~F.~Ginzburg for addressing the question of
the origin of the different azimuthal distributions. 

This work has been supported by the Research Council of Norway.

\end{document}